\documentclass[conference, compsoc, Optimized]{./IEEEtran}
\ifCLASSINFOpdf
\else
\fi
\usepackage{amssymb}
\usepackage{multicol}
\usepackage{boxedminipage}
\usepackage{./zed-csp}
\usepackage{url}
\begin{document}
%
\title{Modelling and Analysing Dynamic Decentralised Systems:\\
 Application to Mobile Ad-hoc Network}


\author{
\IEEEauthorblockN{{Christian Attiogb\'e}
\IEEEauthorblockA{LINA UMR CNRS 6241 - University of Nantes, France\\
Christian.Attiogbe@univ-nantes.fr\\ 
}
}}


%


\maketitle

\begin{abstract}
We introduce a method to specify and analyse decentralised dynamic
 systems; the method is based on the combination of an event-based multi-process
 system specification approach with a multi-facet analysis
 approach that considers a reference abstract model and several specific
 ones derived from the abstract model in order to support facet-wise analysis. 
The method is illustrated with the modelling and the analysis of a mobile
 ad-hoc network. The Event-B framework and its related tools B4free and
 ProB are used to conduct the experiments. \\


\end{abstract}


%
\IEEEpeerreviewmaketitle

\section{Introduction}
\label{section:intro}
%
%
Distributed systems still pose challenging specification and analysis
difficulties that  needs specific languages, methods and tools. 
The structure of a classical centralised software system is based on the composition of
several sub-systems or processes. They are often parallely composed to
enable synchronisation and communication.
Unlikely, decentralised distributed systems with dynamically evolving architecture have unfixed but varying
structure. They cannot be structured with parallel operators that compose
a fixed number of processes; they have an ad-hoc structure related to
the number of involved processes. 
Here, interaction is supported by communication and synchronisation between a
group of processes currently involved in the cooperation to achieve
given goals (the ones defined at the global system level). A group
communication is then needed for systems with dynamic architecture. 

In this article we introduce a \textit{method} for the systematic specification and analysis of
these systems with evolving structure. The proposed method extends and generalises a preliminary work \cite{coloss:CA_ISOLA2008} where a particular approach was used.
We combine a  multi-facet analysis method and a multi-process system specification method
that we developed before \cite{coloss:CA_ICFEM06,coloss:movah08}.
A multi-facet analysis method \cite{AttiogbeSOFSEM05,coloss:CA_ICFEM06,coloss:movah08} consists in studying a system according its various facets both at specification level and verification level. Indeed, it is often the case that a global system is well tackled by combining appropriate languages and techniques that are well-suited for the considered facets of the system.
A multi-process system \cite{coloss:CA_ICFEM06} is the one in which several sub-systems (the processes) are involved, but without a strong composition link between them; thus the architecture of the system is not static, it evolves dynamically according to the existence and the state of the processes.

The contribution of this work is:  a seamless method to globally specify a multi-process system with dynamic architecture by considering an event-based abstract model to guide the specification and by considering several facets during the analysis.

The  article is organised as follows:
in Section \ref{section:manet} we illustrate the features of a decentralised system (MANET) considered as the support of the presentation.
In  Section \ref{section:method} we describe the proposed general modelling
and analysis method.
Section \ref{section:appli2manet} 
presents the application of the proposed method to the MANET system, from the modelling to the  formal analysis.
Finally Section \ref{section:conclusion} concludes the article.

\section{Decentralised Dynamic System}
\label{section:manet}
%
%
%
A Mobile Ad-hoc Network (MANET) system is a typical example of a dynamic decentralised system.
A  Mobile Ad-hoc Network  \cite{Chlamtac2003} is a network formed with wireless mobile
nodes (called ad-hoc nodes) which are the user equipments or devices. 
A MANET has no dedicated network infrastructure, but each node serves as a part of the
network and acts a \textit{router} to forward messages or packets since
there is no router dedicated to that task.

A mobile ad-hoc network is formed only when a group of users
put together their resources to enable and perform communications; hence
a mobile ad-hoc network is dynamically created and may also disappear quickly. 

In a MANET, the nodes communicate either by exchanging
directly or via intermediate nodes. Technically they use ISM
band\footnote{they are radio system frequencies initially dedicated to
industrial, scientific and medical usage.} and
more generally Wireless LAN technologies. Each node is equipped with one
or more radio interfaces with specific transmission features. The
\textit{transmission range} of a node is the transmission area accessible from
this node. All the nodes in this range are accessible directly (one
hop); they are called the neighbours. To address a known node which is not in its transmission range,
the sender node sends its packet to one of the neighbour nodes which is closer
to the destination node (according to the transmission ranges).  
Each node may communicate directly or indirectly using relay nodes
(multi-hop), with other nodes that are outside the sender range.

\medskip
\textit{Dynamic Aspect.} One of the main features of a MANET is its
dynamic aspect:  the structure or topology of the network is frequently
changing. A node may join or leave the net at any time, changing the net
topology. The structure or topology of the net is then highly dynamic. 

\medskip
\textit{Mobility Aspect.} The ad-hoc nodes may move at any time and very frequently due to their mobile nature; 
consequently this impacts not only on the net topology but also on its
quality; there may be route changes, information loss, partitions of the network into different networks, etc. 
As far as routing is concerned, in classical infrastructure-based network,
there are one or several nodes called routers that are in charge of
routing packets between nodes. For this purpose the routers and the nodes are equipped with a routing table
where there is the information about how to join a given destination
node or a network identified with an Internet Address (IP address).

In the scope of MANET, efficient routing protocols development is a challenging concern. 
A message or packet sent to a node reaches it unless the net is partitioned.
 Concerning the time, it is assumed to be discrete and divided into frames.
A node has a set of neighbour nodes during a frame.
During a frame a node may be idle, it also may send messages, receive
messages, forward the received messages.
Before sending a message to a destination, a source node $sn$ which does
not have the destination node address, sends a route request to get this
destination address. The request travels through the net possibly with multi-hop
and reaches the destination which sends back its address. When the address is received
by $sn$ the latter can send its message to the right destination
address.  \\

The study of MANET is an
active and challenging field as this type of network is rapidly
growing and supporting  small and medium size applications
such as mobile services sharing, wireless peer-to-peer systems, etc.
We chose the field of MANET for this work because it is a challenging
field shared by the fields of computer networks and software
engineering. 
From the software system point of view, the MANET system is a
typical decentralised, asynchronous system with dynamically evolving architecture. 
Moreover, its properties (dynamicity, mobility, correctness, etc) need a
combined use of several verification techniques (namely a multifacet analysis).

\section{The Proposed Method}
\label{section:method}

We propose a multi-facet analysis method to globally specify and analyse a
given system using possibly several tools. 
The method is made of four steps as follows:
\vspace{-0.3cm}
\subsection{Overview of the Method}
\begin{description} 
\item[Step 1.]~To build an abstract formal model from the system at hand and
      to state the desired global properties according to this formal
      model; it is the \textit{reference model}; an abstract reference model may be event-based, state-based, process-based, algebraic, etc.
\item[Step 2.]~To systematically derive or translate from this reference model,
      other formal models which are specific to various analysis techniques; 
\item[Step 3.]~To perform analysis (verification of properties) with the specific
      models or with their extensions, by adding specific properties to
      the global ones;
\item [Step 4.]~To ensure the consistency between the reference model and the
      specific ones by propagating the feedback from the specific
      models study on the reference model and by updating consequently
      the other specific models.  Then, the analysis of each facet via a
      specific model participates in the global system analysis.
\end{description}

\medskip
The first step on building a reference model needs methods that are appropriate to the system at hand. In the
current case of the decentralised system, we have a \textit{multi-process} system.  We detail this step.

\subsection{On Building an Event-based Reference Model}
An event-based model is suitable for dynamic system. 
The approach \cite{coloss:CA_ICFEM06} provides rigorous guidelines to help in discovering and expressing the desired
behaviours of a multiprocess system with dynamic architecture. 
Our approach to build the reference model combines a process-oriented view (at low level, for elementary identified processes) and
an event-based one (at global level, for composing processes).
The method used to build the reference model is summarised as follows.
\begin{enumerate}
\item Structuring aspects: From the requirements, elementary \textit{types of processes} are identified to describe behaviours. 
Several processes may have the same type.

Each identified \textit{type of process} $P_i$ that participates in
      the global system model is specified by considering its space state $S_i$
      and the events $E_i$ with their description $Evt_i$ that lead its
      behaviour. 
\centerline{$P_i \defs \langle S_i, E_i, Evt_i\rangle$}
The constituents $S_i$, $E_i$ and $Evt_i$ will be detailed latter on. To handle the dynamic architecture of global system, we impose that for each type of process, the events to \textit{join} and \textit{leave} the system be defined. Some events may be common to several processes; they handle
      interaction and state sharing aspects.
\item Interaction aspects:  Interaction involves communication.  As far as communication is concerned we use
      guarded events, 
      message-passing and ordering of event occurrences. The processes synchronise and communicate through
      the enabling/disabling of the guards of their events. Therefore, if an event
      is used to model a process which is waiting for a data, it may be blocked until the availability of the
      data (enabling the event guard), which is the effect produced by
      another process event. Consider for
      example the case of processes exchanging messages, one process
      waits for the message, hence there should be an event with a non-enabled guard, and another process sends the message via a behaviour of an event which guard is enabled. 
Communications are modelled with abstract channels.
An abstract channel modelled as a set, is used to wait for a message
      or to deposit it. Hence the interaction between the processes is
      handled using  common abstract channels. Therefore, the communication is
      achieved in a completely decoupled way to favour dynamic structuring. 
\item Composition of the processes:
All the described processes are (hierarchically) combined by a fusion operation ($\biguplus$) that merges
      state spaces and the events of the processes into a single global system $S$.
$$ S \defs \biguplus_{i} P_i ~~\equiv~~ \biguplus_{i}~\langle S_i, E_i, Evt_i \rangle$$ 

According to the fusion operator, when the processes are merged, a set is introduced to identify the merged processes. Each feature that is modelled with a variable, results in a function from the set of process identifiers to a set of values (of the feature). The events of the processes are now defined by considering the elements of the identifier set (if the set is empty there is no more process).   
\end{enumerate}

In the following we illustrate  the four steps of the method.

\section{Application to the MANET System}
\label{section:appli2manet}
%
The method is practically supported by tools.
\subsection{The Used Methods and Tools}
\subsubsection{The Event-B Method}
Within the  Event-B framework,  asynchronous systems may be developed and structured using \textit{abstract systems} \cite{Abr96a,AbrialMussat98}. 
\textit{Abstract systems} are the basic  structures of the so-called \textit{event-driven} B. 
An \textit{abstract system} \cite{Abr96a,AbrialMussat98} describes a mathematical model of a system behaviour\footnote{A system behaviour is the set of its possible transitions from state to state beginning from an initial state}.
An abstract system is made mainly of a state description (constants, properties, variables and invariant) and several \textit{event} descriptions. 
Abstract systems are comparable to Action Systems \cite{BaKu83}; they describe a nondeterministic evolution of a system through guarded actions. 
Dynamic constraints can be expressed within abstract systems to specify various liveness properties \cite{AbrialMussat98,ZB02-Cansell}. 
The state of an abstract system is described by variables and constants linked by an invariant. Abstract systems may be refined to concrete ones like abstract machines \cite{ZB02-Cansell,AbrialCansellMery03}. 

An event of a B abstract system is considered as the observation of one transition of the system. Events are  spontaneous and show the way a system evolves. 
An event $e$ is modelled as a \textit{guarded substitution}: $e \defs eG \Longrightarrow eB$ where $eG$ is the event \textit{guard} and $eB$ the event \textit{body} or \textit{action}. 
The symbol $\Longrightarrow$ denotes the guard.\\
 
An event may occur or may be observed only when its guard holds. 
The action of an event describes, with generalised substitutions, how the system state evolves when this event occurs.
Several events may have their guards hold simultaneously; in this case, only one of them  occurs. The system makes internally a nondeterministic choice. If no guard is true the abstract system is blocking (deadlock).
\begin{figure}[htp]
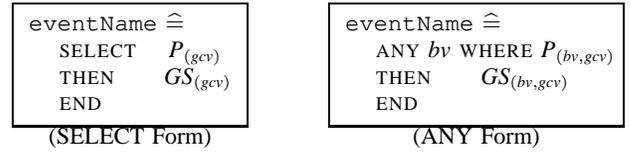

{\small
\begin{center}
\begin{multicols}{2}
\noindent
\begin{boxedminipage}{5.1cm}
\begin{tabbing}
\hspace{0.1cm}\=\hspace{0.4cm}\=\hspace{0.4cm}\=\hspace{0.4cm}\kill
\>\texttt{eventName} $\defs$ \textsf{~~~~~~~} \\
\>\>\textsc{select}~~~
     $P_{(gcv)}$ \\
\>\>\textsc{then} ~~~~
      $GS_{(gcv)}$\\
\>\>\textsc{end}
\end{tabbing}
\end{boxedminipage}
\centerline{(SELECT Form)}
\noindent
\begin{boxedminipage}{4.1cm}
\begin{tabbing}
\hspace{0.1cm}\=\hspace{0.4cm}\=\hspace{0.4cm}\=\hspace{0.4cm}\kill
\>\texttt{eventName} $\defs$ \textsf{~~~~~~ } \\
\>\>\textsc{any} $bv$ \textsc{where}
     $P_{(bv, gcv)}$ \\
\>\>\textsc{then} ~~~~
     $GS_{(bv,gcv)}$\\
\>\>\textsc{end}
\end{tabbing}
\end{boxedminipage}
\centerline{(ANY Form)}
\end{multicols}
\caption{General forms of B events}
\label{figure:eventshape}
\end{center}
}
\end{figure}

An event has one of the general forms (Fig. \ref{figure:eventshape}) where $gcv$ denotes the global constants and variables of the abstract system containing the event; $bv$ denotes the bound variables (variables bound to \textsc{any}). $P_{(bv, gcv)}$ denotes a predicate $P$ expressed with the variables $bv$ and $gcv$; in the same way $GS_{(bv,gcv)}$ is a generalised substitution $S$ which models the event action using the variables $bv$ and $gcv$.
The \textsc{select} form is a particular case of the  \textsc{any} form. The guard of an event with the \textsc{select} form is $P_{(gcv)}$.
The guard of an event with the \textsc{any} form is $\exists(bv).P_{(bv,gcv)}$.
\subsubsection{The B4free and ProB Tools}
We use the theorem prover  
\textsf{B4free}\footnote{{\small\url{www.B4free.fr}}} 
and the \textsf{ProB}\footnote{{\small \url{www.stups.uni-duesseldorf.de/ProB/}}} model checker.

\subsubsection*{Overview of B4free}
\textsf{B4free} is one of the public domain theorem prover  dedicated to Event-B. The prover originated from the industrial commercialised tool called AtelierB. It was developed together with an emacs front-end. It does not have the other modules such as code generator or document generator available in the AtelierB tool.
However, the \textsf{B4free} tool is free and is convenient for experimentations with the B method: parsing and proving the obligation proofs related to the B method. A new public domain framework for B is now available: Rodin\footnote{\url{www.event-b.org/platform.html}}.

\subsubsection*{Overview of ProB}
\label{subsection:ovvProB}
%
%
The ProB tool \cite{LeuschelButler:FME03,LeuschelTurner:ZB05} is an animator and a model checker for  B specifications. 
It supports automated consistency checking of B specifications (an abstract machine or a refinement with its state space, its initialisation  and its operations). The consistency checking is performed on all the reachable states of the machine. ProB also provides a constraint-based checking; with this approach ProB does not explore the state space from the initialisation, it checks whether applying one of the operation can result in an invariant violation independently from the initialisation. 
ProB provides functionalities to show graphical views of automata.
The functionalities of ProB  are organised within three categories: \textit{Animation}, \textit{Verification} and \textit{Analysis}.
ProB tool is used in our study to help in discharging consistency
proof obligations
(invariant violation) and to check liveness properties.

\subsection{Modelling and Analysing the System}
We consider the four steps of the proposed method.
\subsubsection{Step 1: Building an Abstract Reference Model}
The Manet system is made of a set of nodes that communicate; they form a range;  a node is identified as a process type. 
\subsubsection*{Specifying a Node Process}
Each node has
an identifier, a location, an IP address, a connection relation that
indicates its neighbours, etc.  Accordingly we have the $S_i$ part of
the node as a set of typed variables that denote the features.
A set of events ($E_i$) with the associated behaviours ($Evt_i$) define the process behaviours which lead the evolution of the system. As far as its behaviour is concerned, any node
may initiate a message for a given destination, send a message, receive
a message, forward a message, leave a net (a transmission range). 
The behaviour described by these events is observed only when a net exists; that means the net
structuring events are related to those needed for the routing.
Also we deal with the creation of a network by nodes
which have a given range, other nodes may join or leave this
range. Therefore, we link the range of a node with a given abstract
network. 
The formal specification of a MANET  is then 
a set of sequences of configurations of the considered nodes; that is 
their state variables, resulting from the fusion of the node state variables; 
the evolution is modelled through the enabling of
events which possibly modify the state space.
Concerning the interaction within the MANET system, we consider the
events of the nodes and also the common events related to the entire system network (including the ranges). 
\subsubsection{Step 2: Deriving an Event-B Specification}
The derivation of a Event-B model from the abstract event-based model is quite straightforward. The state variables form the B invariant; the abstract events are translated  as B events. 
\subsubsection*{Resulting B specification of the MANET}
The specification of the structure of a MANET is achieved using a set of state variables
and an invariant  that describes the nodes and their current
 configurations:

\begin{center}
{\small
\begin{boxedminipage}{6cm}
\begin{tabbing}
\hspace{0.4cm}\=\hspace{0.4cm}\=\hspace{0.4cm}\=\hspace{0.4cm}\kill
\textbf{\textsc{system}} ${MANET}$\\
\textbf{\textsc{sets}} NODE, RANGE, MSG /* abstract sets */\\
\textbf{\textsc{variables}}\\
\> $nodes, ranges, messages, $ /* state variables*/ \\
\> $rangNodes, reqMsg, inReqMsg, \cdots$\\
 \textsc{invariant} ~~ /* state space predicate */\\
 	\> $nodes \subseteq NODE$ $\land$  $ranges     \subseteq RANGE$\\
$\land$ \> $messages   \subseteq MSG$ \\
$\land$ \>  $rangNodes  \in ranges \rel nodes$\\
$\land$ \> $reqMsg \in nodes \rel messages $\\
$\land$ \> $inReqMsg \in nodes \rel messages$\\
$\land$ \> $inRspMsg \in nodes \rel messages$\\
$\land$ \>  $waitReqMsg \in nodes \rel  messages$\\
$\land$ \> $\cdots$\\
\textbf{\textsc{initialisation}}\\
\> $nodes,ranges, messages,rangNodes := \emptyset,\emptyset,\emptyset,\emptyset $ \\
 $\|$\> $\cdots$ 
\end{tabbing}
\end{boxedminipage}
}
\end{center}

Event-B uses the set notation. The standard operators are written as usually ($\in,~\cup,~\cap,~\{\cdots\}$). The symbol $A \rel B$ denotes a relation between two sets.  

The
 behaviour of the system depends on the set of events that define the
 nodes and the specific system events: the observation of a net creation
 (\textit{newRange}); an
 existing net may disappear if there is no more connected nodes (\textit{rmvRange}).
 Other events considered for the network are the following;
\textit{joinRange}:  a node joins a range;
\textit{leaveRange}: a node leaves a net range;
\textit{newNode}: a new node appears;
\textit{newMsg}: a node initiates a message.

At this stage, the behaviour part (the set of events) is specified using Event-B as follows:\\
\begin{center}
{\small
\begin{boxedminipage}{3cm}
\begin{tabbing}
\hspace{0.4cm}\=\hspace{0.4cm}\=\hspace{0.4cm}\=\hspace{0.4cm}\kill
\textbf{\textsc{system}} ${MANET}$ (continued)\\
\>$\cdots$ \\
\textbf{\textsc{events}}\\
\>\texttt{newNODE} $\defs~~~\cdots$ \\
;\>\texttt{newRANGE} $\defs~~~\cdots$ \\
;\>\texttt{joinRange} $\defs~~~\cdots$ \\
;\>\texttt{leaveRange} $\defs~~~\cdots$ \\
;\>\texttt{newMsg } $\defs~~~\cdots$ 
\end{tabbing}
\vspace{-0.3cm}
\textbf{\textsc{end}}
\end{boxedminipage}
}
\end{center}

The specification of the event \textit{joinRange} is depicted in
Fig. \ref{figure:eventJoinRange}. The other events are specified in
quite the same way.

\begin{figure}[htp]
{\small
\begin{center}
\begin{boxedminipage}{8cm}
\begin{tabbing}
\hspace{0.0cm}\=\hspace{0.4cm}\=\hspace{0.4cm}\=\hspace{0.4cm}\=\hspace{0.9cm}\=\hspace{0.9cm}\kill
\>\texttt{joinRange} $\defs$ ~~ /* a node joins a range */  \\
\>\textsc{ANY} $nd, rg$ \>\>\>\> \textsc{WHERE}\\ 
\>\>	$nd \in nodes \land rg \in ranges $\\
\>\>    $\land rg \in \dom (rangNodes)
	\land  nd \notin rangNodes[\{rg\}]$  \\
\>	\textsc{THEN} ~~\\
\>\>	$rangNodes := rangNodes \cup \{rg \mapsto nd\}$ \\
\>	\textsc{END}
\end{tabbing}
\end{boxedminipage}
\end{center}
}
\caption{{\small Specification of the \texttt{joinRange} event}}
\label{figure:eventJoinRange}
\end{figure}

In the B language, $\dom$ denotes the domain of a relation; if $r$ is a relation and $e$ an element of its domain, $r[\{e\}]$ denotes the images of $e$ with $r$. $a \mapsto b$ denotes the couple $(a, b)$.

As far as the routing aspect is concerned we consider one of the widely studied
routing protocols of MANET: \textit{Ad-hoc On demand Distance Vector}
(AODV) \cite{Chlamtac2003}.  

A part of the behaviour of our B specification is related to
the structuring and another part is about the routing protocol. Therefore we
complete the previous specification of the MANET system with
the events related to the routing protocol.
Within the AODV protocol, each node acts as a router, contributes to
construct routes and to forward messages to other nodes.
There are two phases of the protocol: route discovery and route
maintenance. Route discovery is achieved by exchanging Route Request  (RREQ) and Route Response (RREP) messages.
 The algorithm of the nodes is as follows: when a node desires to set up
 a route to a destination node, it  broadcasts a RREQ message to its
 neighbours (the nodes in its range). The RREQ/RREP messages have the  following main parameters: the source node Id, the destination node Id, the
 number of hops.

 When a node $nd$ receives a RREQ message,   
\textit{i)} either $nd$ is itself a destination and  $nd$ responds with
a RREP 
or $nd$ is an active route to the searched destination node then
$nd$ responds with a route information using the RREP message; 
\textit{ii)} otherwise $nd$ broadcasts the RREQ further with the hop
count of RREQ increased by 1.
The routing of messages is symmetric when a node receives a RREP message. 
The Event B specification is completed with all the events related to
the  routing protocol described above.

We give in the following (see Fig. \ref{figure:eventRREQ}) the specification of the \texttt{sndRREQ} event to illustrate
the specification principle. Here, any node ($sn$) may send a message ($msg$) that it
has already prepared ($msg \in reqMsg[\{sn\}]$) to all the nodes in its
range ($otherNodesInRange$). Exchanged messages are modelled using
abstract channels ($inRepMsg$,$repMsg$).\\
\vspace{-0.2cm}
\begin{figure}[htp]
{\small
\begin{center}
\begin{boxedminipage}{6cm}
\begin{tabbing}
\hspace{0.0cm}\=\hspace{0.4cm}\=\hspace{0.4cm}\=\hspace{0.4cm}\=\hspace{0.9cm}\=\hspace{0.9cm}\kill
\>\texttt{sndRREQ} $\defs$ ~~ /* route request from sn to dn */  \\
\>\textsc{ANY} $sn, msg$ \>\>\>\> \textsc{WHERE}\\ 
\>\>	$sn \in nodes$  /* source */  \\
\>\>	$\land  msg \in MSG \land msg \in messages$   \\
\>\>	$\land  msg \in reqMsg[\{sn\}]$ /* a msg initiated by nd */  \\
\>	\textsc{THEN} \\
\>\>	\textsc{LET} $otherNodesInRange$ \\
\>\>	\textsc{BE} $otherNodesInRange = \{ndi | ndi \in nodes$ \\
\>\>	$\land ndi \neq sn \land  rangNodes^{-1}(sn) = rangNodes^{-1}(ndi) \}$ \\
\>\>	\textsc{IN} 
	     \>\>$inReqMsg := $\\
\>\>\>\>$~ ~ ~ ~ ~ inReqMsg \cup (otherNodesInRange *\{msg\})$ \\
\>\>\>	$\|$ \>$reqMsg := reqMsg - \{(sn \mapsto msg)\}$  \\
\>\>	\textsc{END} \\
\>	\textsc{END}
\end{tabbing}
\end{boxedminipage}
\end{center}
}
\caption{{\small Specification of the \texttt{sndRREQ} event}}
\label{figure:eventRREQ}
\end{figure}

The expression $otherNodesInRange *\{msg\}$ denotes the Cartesian product of the elements in $otherNodesInRange$ with the singleton $\{msg\}$.

The dynamic aspect of the system architecture is based on the
fact that the event guards depend on the variables $nodes, messages,
\cdots$ which themselves depend on the current event. That means in an event guard, we can consider any event from $nodes$ or any messages from $messages$, etc. 

This is illustrated by
the non-deterministic form of the event specifications:\\
\centerline{\texttt{event} $\defs$ \textsc{ANY} $sn$ \textsc{WHERE} $sn
\in nodes$ \textsc{THEN}  ...	\textsc{END}}

The Event-B specification which is the specific  model for the study of 
the MANET is then an abstract system equipped with
all the events described before: structuring and routing events.\\

\subsubsection{Step 3: Analysis of the Specific Model}

A multi-facet analysis of the specific Event-B model of
the MANET system is performed. For this purpose two different tools are used but they cover different facets of the analysis:
B4free and ProB. 

\subsubsection*{Consistency and Refinement of the System}
The previously described B abstract system is proved consistent using the
B4free tool. Then it is refined; more details are added to the state
space and the event specifications; for instance we consider the management of
the IP addresses of the nodes and  exchanged messages.
Unlike in the abstract system where a packet destination is
nondeterministically selected, in the refinement the nodes and the
messages have IP addresses, therefore, the receiver node is checked
against the destination IP address. 
The resulting refined system is also proved correct with respect to consistency using
the B4free tool. However to accomplish the proofs, we combine the use of
B4free and ProB. That is, when a proof
obligation is not discharged by B4free, we  model-check the
specification and discover possible errors by displaying and analysing
the displayed error
state. Accordingly the feedback is propagated in the reference model and we iterate. 

\subsubsection*{Liveness Properties Analysis}
Many properties of the MANET routing protocol are well-expressed using LTL (Linear Temporal Logic) formula which
is not supported by  the B4free tool. 
We express these liveness properties with the ProB LTL formalism. 
Then we extend
the Event-B abstract system with these LTL properties; for example

\noindent 
{\small \textbf{P$_1$}. \textsf{A route request is always followed by a response:}}\\
\centerline{\small $G(e(sndRREQ) \implies F(e(sndRREP)))$}

The resulting specification is model-checked using ProB.

After that we come to the conclusion that our model  extended with the stated properties, is correct with respect to these properties. 

\subsubsection{Step 4: Feedback to the Reference Model}
Using the multi-facet approach, with B4free and ProB 
 helps us to perform a complete analysis. 
For illustration, in experiments with ProB, when a deadlock is
detected after the exploration of nodes and transitions that cover all the operations (the B events), the state corresponding to the deadlock is carefully analysed. In one case we discover that it corresponds to a situation (net partitioning) where there are nodes with
some packets to be transmitted but no node in the current net range. This
corresponds to a real-life situation which is due to the  dynamic
aspect of the MANET and the mobility of nodes. 
A feedback is then propagated first in the Event-B specification.
The model is corrected  by strengthening the guard of message initiation by the hypothesis of
non-emptiness of the net range. Thus the analysis of the model runs without
errors\footnote{the experiment result tables, not displayed here, show 0
deadlocked states for hundreds of explored states and transitions.}.

\section{Discussion and Conclusion}
\label{section:conclusion}
%
%

We proposed a method that combines two main techniques to
model and analyse dynamic decentralised  systems: a multi-facet analysis technique and an
event-based technique. 
As illustration the MANET system was modelled and analysed using Event-B tools for experimentations.

The proposed method  is to be contrasted with classical approaches (Transition Systems, Process Algebra) where the dynamic aspect of the system is not taken into account;
most of them use a process-algebra oriented approach, they focus on the changes on defined architectures and
(pre)define rules to perform reconfiguration of the architecture. 
Our event-based approach overcomes these limitations of the classical approaches; it  considers distribution and mobility of processes
and no predefined reconfiguration rules are needed, instead we use the
behaviour of process types and the composition of events with related
guards that depend, at the abstract level, on shared state
information.  The $\pi-$calculus
\cite{Milner92} permits the description of evolving structures of
processes but new processes are generated from existing ones with the
name passing mechanisms; the $\pi-$calculus is also not yet well supported by tools.

Ongoing works are about the scalability of our approach; we also plan
a refinement of the specification until simulation (that will replace
code level). Currently, to tackle the scalability we consider the
strengthening of message passing aspects during the refinement of our
specifications; indeed message passing is the standard way to deal
with asynchronous communication. It will be very interesting to get an
abstract specification level where these communications are expressed
with very simple schemas.

\vspace{-0.2cm}
\small
\bibliographystyle{abbrv}


\end{document}